\documentclass[aps,pra,twocolumn,notitlepage,showkeys,nofootinbib,showpacs,superscriptaddress,10pt,amssymb,amsmath]{revtex4-1}
\usepackage{graphicx}
\usepackage{calc}
\usepackage[pdftex]{xcolor}
\definecolor{darkblue}{rgb}{0.0,0.0,0.6}
\usepackage{braket}
\usepackage[pdftex,
        colorlinks=true,
        linkcolor=darkblue,
        citecolor=darkblue,
        filecolor=black,
        urlcolor=darkblue,
        bookmarks=true,
        bookmarksopen=true,
        bookmarksopenlevel=3,
        plainpages=false,
        pdfpagelabels=true,
        verbose=true]{hyperref}
\usepackage[all]{hypcap}

\usepackage{ulem}

\begin{document}

\title{Symmetry-broken momentum distributions induced by matter-wave diffraction during time-of-flight expansion of ultracold atoms}

\author{M.\,Weinberg}
\author{O.\,J\"{u}rgensen}
\altaffiliation{M.\,W. and O.\,J. contributed equally to this work}
\affiliation{Institut f\"{u}r Laserphysik, Universit\"{a}t Hamburg, Luruper Chaussee 149, D-22761 Hamburg, Germany}
\author{C.\,\"{O}lschl\"{a}ger}
\affiliation{Institut f\"{u}r Laserphysik, Universit\"{a}t Hamburg, Luruper Chaussee 149, D-22761 Hamburg, Germany}
\author{D.-S.\,L\"{u}hmann}
\affiliation{Institut f\"{u}r Laserphysik, Universit\"{a}t Hamburg, Luruper Chaussee 149, D-22761 Hamburg, Germany}
\author{K.\,Sengstock}
\email{sengstock@physnet.uni-hamburg.de}
\affiliation{Institut f\"{u}r Laserphysik, Universit\"{a}t Hamburg, Luruper Chaussee 149, D-22761 Hamburg, Germany}
\affiliation{\mbox{Zentrum  f\"ur Optische Quantentechnologien, Universit\"at Hamburg, Luruper Chaussee 149, D-22761 Hamburg, Germany}}
\author{J.\,Simonet}
\affiliation{Institut f\"{u}r Laserphysik, Universit\"{a}t Hamburg, Luruper Chaussee 149, D-22761 Hamburg, Germany}
\affiliation{\mbox{Zentrum f\"ur Optische Quantentechnologien, Universit\"at Hamburg, Luruper Chaussee 149, D-22761 Hamburg, Germany}}

\begin{abstract}
We study several effects which lead to symmetry-broken momentum distributions of quantum gases released from optical lattices. In particular, we demonstrate that interaction within the first milliseconds of the time-of-flight expansion can strongly alter the measurement of the initial atomic momentum distribution. For bosonic mixtures in state-dependent lattices, inter-species scattering processes lead to a symmetry breaking in momentum space. The underlying mechanism is identified to be diffraction of the matter wave from the total density lattice, which gives rise to a time-dependent interaction potential.
Our findings are of fundamental relevance for the interpretation of time-of-flight measurements and for the study of exotic quantum phases such as the twisted superfluid. Beyond that, the observed matter-wave diffraction can also be used as an interferometric probe.
In addition, we report on diffraction from the state-dependent standing light field, which leads to the same symmetry-broken momentum distributions, even for single component condensates.
\end{abstract}
\pacs{03.75.Lm, 03.75.Mn, 37.10.Jk, 67.85.Hj}
\maketitle

\section{Introduction}\vspace{-3mm}

Ultracold atoms in optical lattices developed to an intense field of research for new quantum many-body systems as well as model systems with strong connections to condensed matter physics \cite{Bloch:2008gl,Lewenstein:WpY7sZa0}.
Optical lattices themselves constitute a very powerful and versatile tool: the intensity modulation resulting from the interference of far-detuned laser beams allows for various lattice geometries as for instance triangular, honeycomb or kagome \cite{Grynberg:1993bm,Becker:2010de,Soltan-Panahi:2011ey,Tarruell:2012db,Jo:2012br}.
In addition, a selective trapping of specific internal states of one atomic species can be achieved taking advantage of the polarization modulation of the light field. In these so-called \textit{state-dependent} lattices controlled collisions \cite{Mandel:2003fj}, quantum walks \cite{Karski:2009} or 2D arrays of double wells \cite{Lee:2007} have been investigated, to name only a few examples. A pure polarization modulation of the light field realizes exotic mixtures where only one of two constituents is trapped by the lattice potential \cite{McKay:2010jn}.

\begin{figure}[b]
	\centering
	\includegraphics{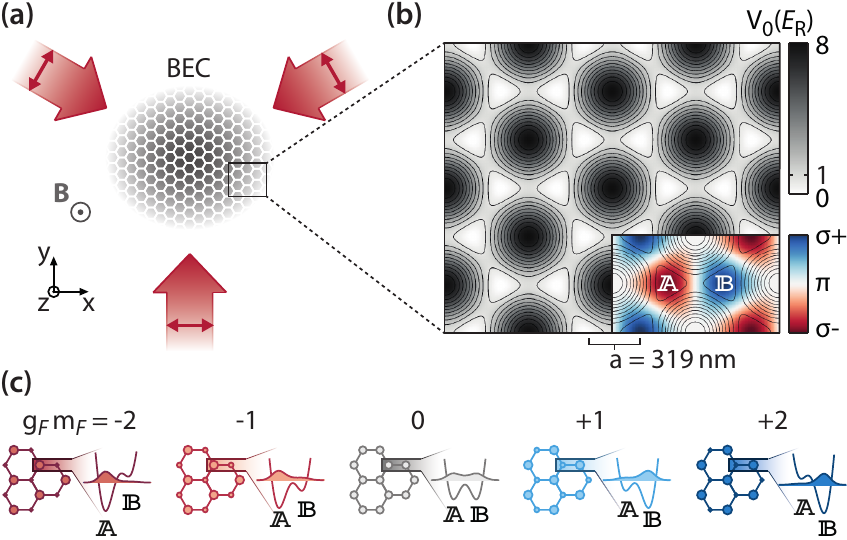}
	\caption[Honeycomb potential]{State-dependent optical honeycomb potential. (a) Illustration of the three-beam setup for the generation of the state-dependent honeycomb potential. All three beams are linearly polarized along the lattice plane. The quantization axis is defined by a magnetic field $\mathbf{B}$ perpendicular to the lattice plane. (b) Resulting potential for an atom with $m_F=0$ and alternating circular polarization pattern (inset). (c) Resulting lattice geometries for atoms in the state-dependent lattice for different hyperfine states of $^{87}\text{Rb}$. Atoms with non-vanishing magnetic quantum numbers are trapped in one of the two triangular sublattices denoted $\mathbb{A}$ and $\mathbb{B}$.}
	\label{fig:01}
\end{figure}

While single-site and single-atom resolved detection is now possible in optical lattices \cite{Bakr:2009bx,Sherson:2010hg,Haller:2015,Cheuk:2015,Parsons:2015}, most experiments to date rely on measurements after time-of-flight. During time-of-flight the atomic ensemble undergoes a ballistic expansion which allows retrieving its momentum distribution. When the coherence length is large compared to the lattice spacing, the momentum distribution of ultracold bosonic atoms is expected to show a sharp interference pattern with the same symmetry as the reciprocal lattice \cite{Greiner:2001cm,Pedri:2001jn}.

\begin{figure*}[!t]
	\centering
	\includegraphics{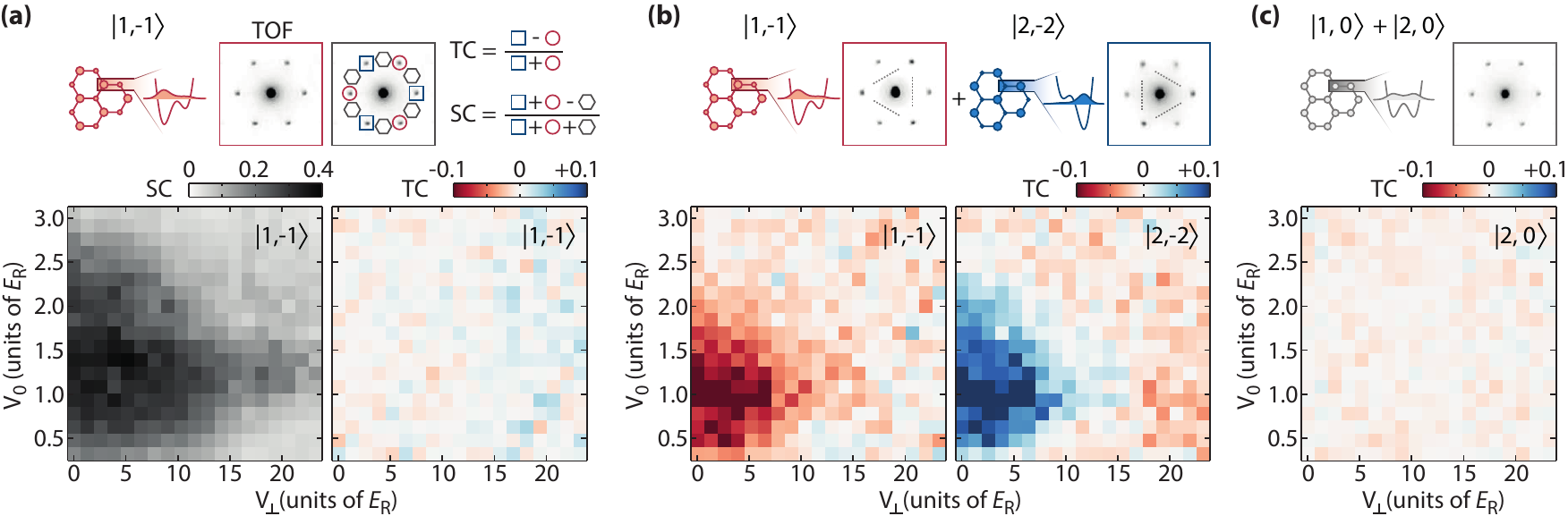}
	\caption[Symmetry breaking in momentum space]{Symmetry breaking in momentum space. (a) Momentum distribution of $\ket{1,-1}$ atoms in the state-dependent honeycomb lattice retrieved from resonant absorption imaging. The superfluid (SC) and triangular (TC) contrast, which quantify respectively the coherence of the superfluid and the reduced three-fold symmetry of the momentum distribution, are reported for different lattice depths $V_0$ and $V_\perp$. (b) For a 1:1 mixture of $\ket{1,-1}$ and $\ket{2,-2}$ atoms, the momentum distribution of each hyperfine state shows a reduced three-fold symmetry. The triangular contrast, opposite for the two states, is reported for different lattice depths $V_0$ and $V_\perp$. (c) For a 1:1 mixture of $\ket{1,0}$ and $\ket{2,0}$ atoms, the momentum distribution of the $\ket{2,0}$ atoms is six-fold symmetric as the triangular contrast vanishes for all lattice depths.}
	\label{fig:02}
\end{figure*}

In this paper we discuss different mechanisms leading to momentum distributions that break the symmetry of the reciprocal lattice. This symmetry breaking can either indicate exotic quantum phases with a complex order parameter \cite{Soltan-Panahi:2011} or result from redistribution processes between different momentum states. As a central result we demonstrate that scattering processes during the time-of-flight expansion can alter the initial momentum distribution much stronger than anticipated so far. Our experimental studies are performed in a state-dependent honeycomb optical lattice \cite{Soltan-Panahi:2011ey} but are relevant for all lattices with multi-atomic basis.
Scattering processes due to the short-range atomic interactions are usually neglected for the description of the time-of-flight expansion \cite{Gerbier:2008bs}. 
However as reported in Ref.~\cite{Pertot:2010}, such processes can give rise to four-wave mixing of a two-component matter wave corresponding to a large redistribution between momentum states. In optical lattices, as the scattering processes have the same symmetry as the reciprocal lattice, they average out and the initial momentum distribution remains unaffected. 

In section \ref{sec2}, we demonstrate that this situation changes drastically for two-component bosonic atoms localized on different sublattices. For a state-dependent optical honeycomb lattice, the momentum distribution measured after time-of-flight breaks the inversion symmetry of the reciprocal lattice. This striking effect results from a redistribution of the atoms between different momentum states induced by inter-species scattering events during the first milliseconds of the time-of-flight expansion. The relevance of these results for the observation of the twisted superfluid phase is discussed in section \ref{sec3}.

\begin{figure*}[!t]
	\centering
	\includegraphics{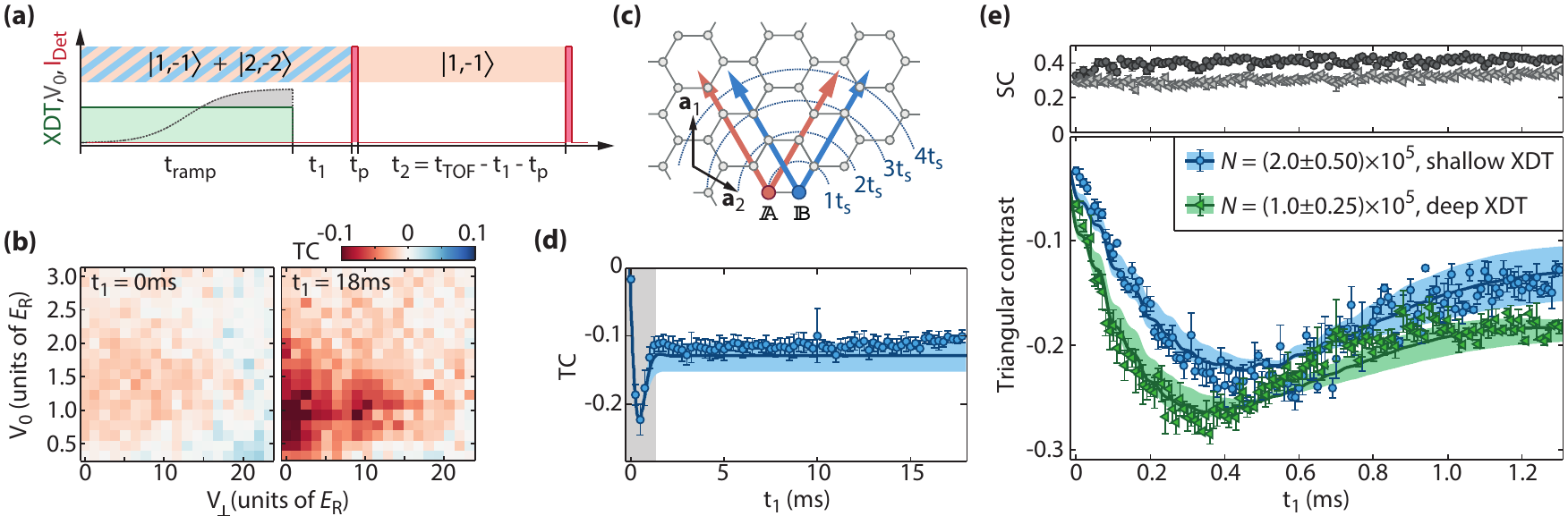}
	\caption[Interaction in TOF]{Interactions in time-of-flight. (a) Experimental procedure. The 1:1 mixture of $\ket{1,-1}$ and $\ket{2,-2}$ atoms is adiabatically loaded in a three-dimensional lattice with final potential depths $V_0$ and $V_\perp$. A resonant light pulse of $t_p=100\,\mathrm{\mu s}$ expels the $\ket{2,-2}$ at a time $t_1$ during the time-of-flight expansion. The momentum distribution of the remaining $\ket{1,-1}$ is imaged after a time-of-flight of $t_\mathrm{TOF}\,=\,27\,\mathrm{ms}$. (b) Triangular contrast measured for $t_1=0\,\mathrm{ms}$ (left) and 18\,ms (right) for different lattice depths $V_0$ and $V_\perp$. (c) Classical model depicting the scattering processes for two different reciprocal lattice vectors, where $t_{\text{s}}$ denotes the time before the first scattering event takes place. (d) Triangular contrast of the momentum distribution as a function of the expelling time $t_1$ for  fixed lattice depths $V_0=0.91 E_R$ and $V_\perp=0 E_R$. (e) Evolution of superfluid (top panel) and triangular contrast (bottom panel) during the first millisecond of the time-of-flight expansion  [corresponding to the gray shaded area in (d)]. Data is shown for a shallow crossed dipole trap (XDT) with trap frequencies of $\nu_{\text{XDT}}=\left( 48,20,20 \right)\,\mathrm{Hz}$ with $N=\left(2.00\pm0.50\right){\times} 10^5$ atoms (circles) and and a deep dipole trap at $\nu_{\text{XDT}}=\left( 120,50,50 \right)\,\mathrm{Hz}$ with $N=\left(1.00\pm0.25\right){\times} 10^5$ atoms (triangles). Error bars represent the standard error of the mean of at least three observations. The numerical results (thick solid lines and filled areas) shown in (d) and (e) exhibit good agreement with the experimental data.}
	\label{fig:03}
\end{figure*}

Such a redistribution between different momenta is also known to occur when a matter wave diffracts from a standing light field \cite{Gould:1986}. As reported in section \ref{sec4}, this \textit{single-particle} effect may as well give rise to symmetry-broken momentum distributions in the case of a state-dependent light potential.

\section{\label{sec1}State-dependent honeycomb potential}
The state-dependent honeycomb lattice is generated by three running-wave laser beams that intersect in the $xy$-plane under angles of $120^\circ$ with an in-plane alignment of the linear polarization vectors as illustrated in Figure\,\ref{fig:01}(a) \cite{Grynberg:1993bm,Becker:2010de,Luehmann2014hex}. This orientation of the polarization vectors gives rise to an alternating pattern of circular polarization in the resulting light field [see Fig.\,\ref{fig:01}(b)] which stems from the projection of the oscillating electric field onto the quantization axis, defined by a homogeneous magnetic field along $z$. For $^{87}\text{Rb}$ atoms and a laser wavelength of ${\lambda_L}\,{=}\,{830\,}$nm, used in the experiment, the detuning relative to the atomic transitions still is on the order of the fine structure splitting. In addition to the intensity modulation of the resulting light field, this circumstance gives rise to a reasonably strong polarization-induced Stark shift of the magnetic Zeeman substates $\ket{F,m_F}$. The total optical potential can be expressed as a sum of a state-independent $V_{\text{int}}(\mathbf{r})$ and a state-dependent part $V_{\text{pol}}(\mathbf{r})$:
\begin{equation}
	V(\mathbf{r})= -V_0\Big[V_{\text{int}}(\mathbf{r}) + V_{\text{pol}}(\mathbf{r})\Big].
\end{equation}
Here, $V_0$ denotes the corresponding lattice depth created for two equivalent counter-propagating laser beams (see Appendix\,\ref{AppendixA} for more details). It is commonly given in units of the recoil energy $E_R=h^2/(2M\lambda_L^2)$, where $h$ is the Planck constant and $M$ the atomic mass of $^{87}$Rb.

Hence, for magnetic quantum numbers different from zero, the alternating pattern of circular polarizations breaks the inversion symmetry of the honeycomb lattice: the potential energy of one of the two triangular sublattices $\mathbb{A}$ and $\mathbb{B}$ is lifted while the other one is lowered [compare Fig.\,\ref{fig:01}(c)]. Whether an atom in a hyperfine state $\ket{F,m_F}$ is predominantly confined at a lattice site with $\sigma^+$ or $\sigma^-$ polarization depends on the sign of its magnetic quantum number and the respective Land\'e factor $g_F$. Accordingly, an atom in the hyperfine state $\ket{1,-1}$ experiences the same potential as an atom in the state $\ket{2,+1}$ and both are mainly trapped in the same sublattice.

\section{\label{sec2}Impact of scattering in Time-of-Flight}
In this section, we report on the symmetry breaking in momentum space occurring for specific mixtures of hyperfine states in the state-dependent honeycomb lattice. For quantum gas experiments, it is usually assumed that the first order correlations determine the structure factor of the lattice, which in turn determines the atomic momentum distribution after time-of-flight \cite{Greiner:2001cm,Pedri:2001jn}. Here we demonstrate that this mapping has to be revised for lattices with multi-atomic basis as scattering processes during the early stages of time-of-flight expansion may strongly alter the outcome of such observations.

As shown in Figure\,\ref{fig:02}(a) for atoms in the magnetic hyperfine state $\ket{1,-1}$, all first-order interference peaks have the same amplitude. Indeed, when the coherence length is large compared to the lattice spacing, the momentum distribution shows a sharp interference pattern with the same six-fold rotational symmetry as the reciprocal lattice. The interference pattern is often characterized by its visibility or superfluid contrast, which is reported in Figure\,\ref{fig:02}(a) as a function of the potential depths $V_0$ and $V_\perp$ of the state-dependent honeycomb lattice and an additional 1D lattice respectively.


For mixtures of different hyperfine states, the momentum distribution observed after time-of-flight reveals new features, which strongly depend on the magnetic states involved. Figure\,\ref{fig:02}(b) shows the momentum distributions for a 1:1 mixture of $\ket{1,-1}$ and $\ket{2,-2}$ atoms, which are localized on the two complementary triangular sublattices $\mathbb{A}$ and $\mathbb{B}$ respectively. These momentum distributions, obtained after applying a Stern-Gerlach field in order to separate the two components, reveal a striking symmetry breaking in momentum space, which appears as an alternating pattern in the first-order diffraction peaks. A quantitative measure of this reduced three-fold rotational symmetry is given by the triangular contrast [see Fig.\,\ref{fig:02}(a)], which is opposite for the two hyperfine states at hand. This triangular contrast has been systematically measured for different lattice depths $V_\perp$ and $V_0$ as reported in Figure\,\ref{fig:02}(b). The observed symmetry breaking is a very robust effect occurring for lattice depths at which the superfluid contrast of the momentum distribution is non-zero.
However for a 1:1 mixture of non-magnetic hyperfine states $\ket{1,0}$ and $\ket{2,0}$, delocalized on a honeycomb lattice, the six fold symmetry of the momentum distribution is preserved as the triangular contrast vanishes for all lattice depths $V_\perp$ and $V_0$ [compare Fig.\,\ref{fig:02}(c)]. Note that here, only the $\ket{2,0}$ atoms have been imaged by omitting the repumping light as a separation by a Stern-Gerlach field is not possible.

\begin{figure}[t]
	\centering
	\includegraphics{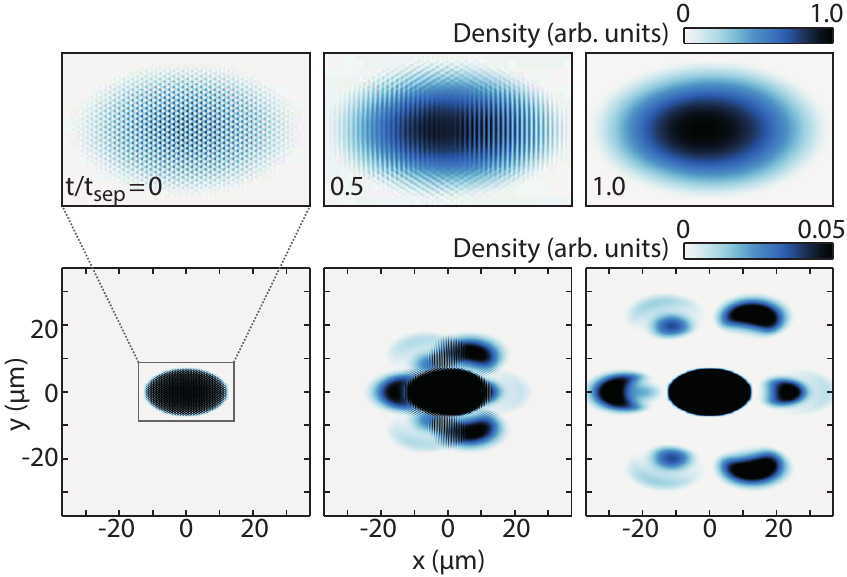}
	\caption[ToF]{Matter-wave diffraction in time-of-flight. Calculated density distribution of the $\ket{1,-1}$ atoms obtained by solving the time-dependent Gross-Pitaevskii equation during the time-of-flight expansion in presence of the $\ket{2,-2}$ atoms for three different times $t_1/t_\mathrm{sep}=0$, $0.5$ and $1.0$ (bottom panel) in units of the separation time $t_\mathrm{sep}$ after which the atomic clouds corresponding to different momentum states are separated. The re-scaled central densities are shown in inset (top panel).}
	\label{fig:04}
\end{figure}

As we show in the following, this symmetry breaking in momentum space results from inter-species scattering occurring at the beginning of the time-of-flight expansion. Hyperfine states with opposite $g_Fm_F$ populate different sublattices of the state-dependent honeycomb lattice and their density distribution forms an interaction induced lattice from which they diffract. The scattering processes for each atom strongly depend on its momentum as illustrated in Figure\,\ref{fig:03}(c): after a fixed expansion time, an atom on sublattice $\mathbb{A}$ (red) may have encountered up to 1/3 more lattice sites along the reciprocal vector $b_1$ than along $b_2$. For atoms localized on the sublattice $\mathbb{B}$ (blue), this effect is inverted. This simplified picture evidences how momentum-dependent scattering leads to a redistribution of the atoms between the different diffracted orders and induces the triangular contrast in the momentum distribution. This redistribution decays in time as the diffracted atomic clouds separate spatially.  This occurs within the separation time $t_\mathrm{sep}\approx \,2.7\,\mathrm{ms}$ for our parameters (see Appendix\,\ref{AppendixB}).

To investigate the impact of the inter-species scattering on the mixture of $\ket{1,-1}$ and $\ket{2,-2}$ atoms, the $\ket{2,-2}$ component is removed at different times $t_1$ [see Fig.\,\ref{fig:03}(a)]. The triangular contrast of the remaining species is again analyzed for different lattice depths $V_\perp$ and $V_0$ as depicted in Figure\,\ref{fig:03}(b).
The symmetry breaking only appears if the second component is present during time-of-flight. The contrast then increases with the time the second component is present during time-of-flight. Indeed, when the resonant light pulse is applied at the start of the time-of-flight expansion ($t_1$\,=\,0\,ms), the six-fold symmetry in momentum space is restored. For $t_1$\,=\,18\,ms, a non-vanishing triangular contrast is measured, in agreement with the data presented in Figure\,\ref{fig:02}(b).

As reported in Figure\,\ref{fig:03}(d) the triangular contrast of the $\ket{1,-1}$ atoms indeed increases from zero to its final amplitude within the first 1.2\,ms of the time-of-flight expansion due to the presence of the $\ket{2,-2}$ atoms with a pronounced maximum at almost double the final value. The time evolution of the superfluid and triangular contrast during the first milliseconds of the time-of-flight expansion has been measured for different atomic densities obtained by varying the strength of the overall harmonic confinement as shown in Figure\,\ref{fig:03}(e). Experimental data of the triangular contrast are in excellent agreement with numerical simulations of the time-dependent Gross-Pitaevskii equation describing the time-of-flight expansion in analogy to Ref.\,\cite{Pertot:2010} (see Appendix\,\ref{AppendixB}). Contrary to the behavior of the triangular contrast, the superfluid contrast remains at a constant value which rules out possible heating processes induced by the resonant light pulse used to remove the second component.

Figure\,\ref{fig:04} shows the calculated evolution of the density distribution of the $\ket{1,-1}$ atoms at the beginning of the time-of-flight expansion. Atoms are redistributed among the different momentum states by diffraction on the interaction induced lattice. In addition to the emerging triangular contrast of the diffraction peaks, the structure of the diffraction peaks itself is strongly altered by the scattering processes and displays in space the oscillatory behavior observed for the triangular contrast over time. Together with the initially elliptic density profile of the atomic cloud that is elongated along the $x$ direction the resulting density distribution is symmetric only with respect to the $x$-axis. 

\begin{figure}[t]
	\centering
	\includegraphics{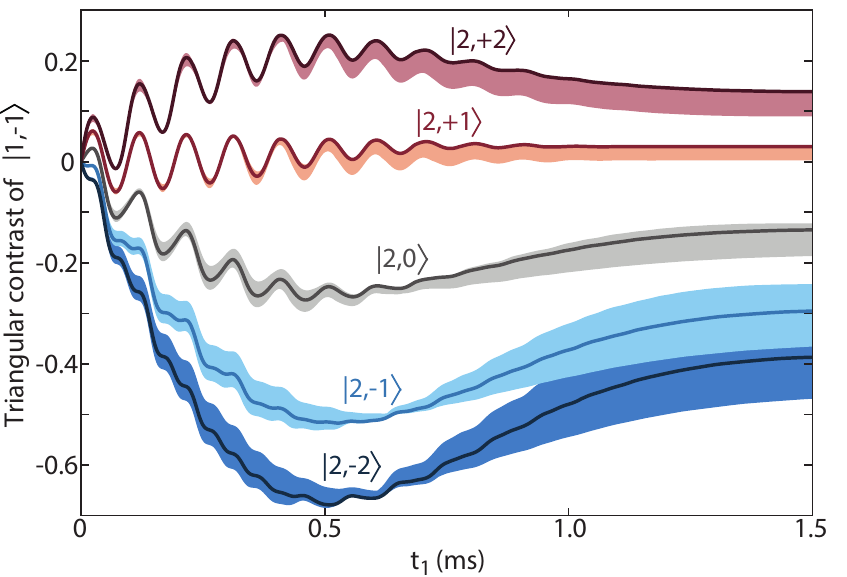}
	\caption[Mixtures]{Calculated evolution of the triangular contrast in dependence of the expansion time $t_1$ for $1{:}1$ mixtures of $\ket{1,-1}$ with different $\ket{2,m_F}$ states as shown in the plot. Solid lines and filled areas correspond to calculations for particle numbers of $N=\left(2.00\pm0.5\right){\times}10^5$ at lattice depths of $V_0=0.91 E_R$ and $V_\perp=0 E_R$.}
	\label{fig:05}
\end{figure}

Beyond the specific case of a mixture of $\ket{1,-1}$ and $\ket{2,-2}$ atoms, numerical results of the triangular contrast of the $\ket{1,-1}$ state with respect to the expansion time $t_1$ are shown for various other $1{:}1$ mixtures of $\ket{1,-1}$ with different $\ket{2,m_F}$ states in Fig.\,\ref{fig:05} [note that the results for the $\ket{2,-2}$ component are the same as shown in Fig.\,\ref{fig:03}(d) and \ref{fig:03}(e) without re-scaling of the total amplitude]. The strong oscillatory behavior of the triangular contrast that can be observed for mixtures with $\ket{2,0}$, $\ket{2,+1}$, and $\ket{2,+2}$ components stems from the re-distribution of atoms from $\mathbb{A}$ to $\mathbb{B}$ sites (and vice versa) on the timescale of the scattering time $t_{\text{s}}$ [compare Fig.\,\ref{fig:03}(c)]. Remarkably, even mixtures that are predominantly located at the same sublattice exhibit distinct dynamics of the triangular contrast. Moreover, the $\ket{1,-1}$ and $\ket{2,+1}$ mixture in fact represents the behavior of a single-species since the potential is the same for both components. While the resulting amplitude of the triangular contrast is comparatively small, this observation implies that matter-wave diffraction in the early stages of time-of-flight expansion cannot be neglected even for single-component experiments trapped in lattice structures with multi-atomic basis.

In this section, we have shown that for bosonic mixtures interaction strongly alters the momentum distribution measured after time-of-flight. The localization of the two components on different sublattices induces an asymmetry in the scattering processes occurring during the first millisecond of the time-of-flight expansion, yielding a symmetry-broken momentum distribution. Strikingly, each atom scatters at least once during the expansion. However, it is still possible to retrieve the unperturbed momentum distribution if one of the two components is removed right before the expansion starts or by eliminating interactions by using Feshbach resonances.

\section{\label{sec3}Twisted superfluid phase}
A symmetry breaking in momentum space can also be attributed to a complex-valued superfluid order parameter as realized in a twisted superfluid phase \cite{Soltan-Panahi:2011}. Thereby the phase of the superfluid order parameter is twisting between neighboring sites $i$ and $j$ as illustrated in Figure\,\ref{fig:04}(a), causing a symmetry-broken momentum distribution for this state in the optical lattice \cite{Jurgensen:2015}. For two components, opposite triangular pattern in momentum space are expected due the opposite orientation of the complex order parameter ($-\theta$ and $\theta$).

\begin{figure}[t]
	\centering
	\includegraphics{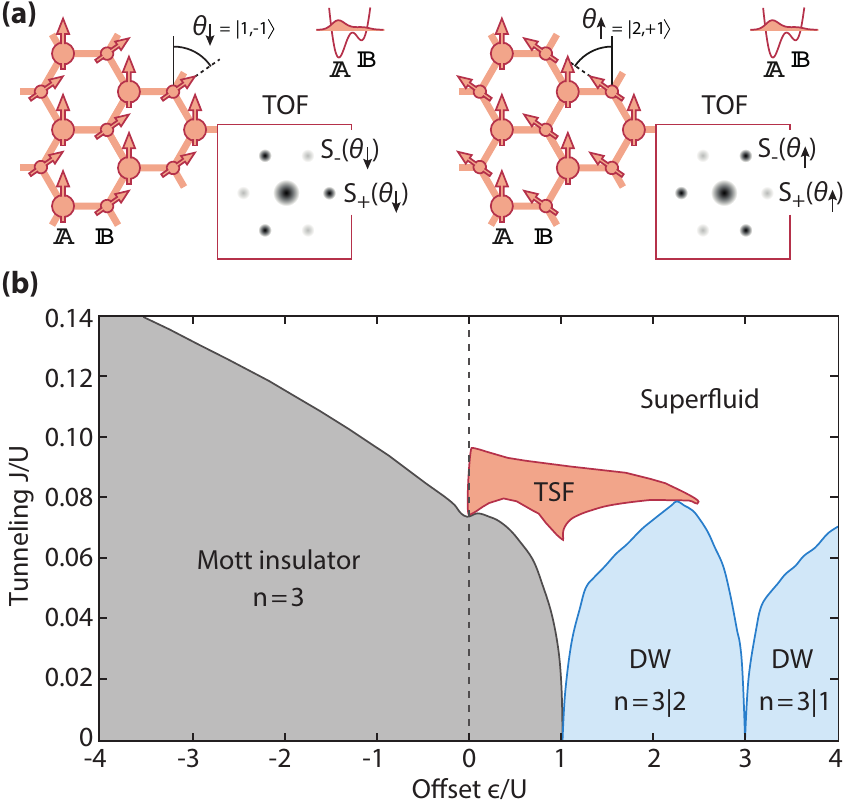}
	\caption[TSF]{Twisted superfluid phase. (a) The complex phase of the superfluid order parameter is twisting between neighboring sites leading to a reduced three-fold symmetry for the momentum distribution shown in the inset. For two components, the opposite orientation of the complex order parameter ($-\theta$ and $\theta$) results in opposite triangular patterns in momentum space. (b) Phase diagram for a two-component bosonic superfluid in a state-dependent honeycomb lattice as a function of the tunneling amplitude ($J/U$) and the energy offset between the sublattices $\mathbb{A}$ and $\mathbb{B}$ ($\epsilon/U$). The twisted-superfluid phase (TSF) is predicted only for small energy offsets, while for larger energy offsets the superfluid phase is favoured. Depending on the offset, the insulating phases are Mott insulators or density waves (DW).}
	\label{fig:06}
\end{figure}

However, in order to retrieve the momentum distribution of the twisted superfluid phase, the time-of-flight expansion should not be affected by interaction effects as described in the previous section. In the measurements presented in Figure\,\ref{fig:03}(b), where one component is removed at the beginning of the time-of-flight, the vanishing triangular contrast indicates a normal superfluid phase in the investigated parameter space. Furthermore, the observation of a specific orientation of the triangular pattern for each species reported in \cite{Soltan-Panahi:2011} can also be explained by scattering between the two species during time-of-flight. So far, this was standing in contradiction to the theoretical description as the twisted superfluid phase is two-fold degenerate associated with a spontaneous breaking of the time-reversal symmetry.

The emergence of this twisted superfluid phase was first attributed to pair processes in an effective single-component mean-field picture \cite{Soltan-Panahi:2011}. Subsequent theoretical two-component studies have found only real-valued superfluids, but either discard correlations \cite{Choudhury:2013cn} or are restricted to the standard Bose-Hubbard model dismissing pair-tunneling \cite{Cao:2015}. In Ref.~\cite{Jurgensen:2015}, the emergence of the twisted superfluid state was found to be induced by pair tunneling and counter-hopping processes using an extended Hubbard model and a correlated description. In general, the twisted superfluid phase is found for both single- and two-component systems.

Although the two-component twisted superfluid emerges for relatively small pair-tunneling amplitudes, these can only be achieved for scattering lengths several times larger than the one of $^{87}$Rb \cite{Jurgensen:2015}. In addition, state-dependent lattices with a large energy offset between the sublattices $\mathbb{A}$ and $\mathbb{B}$ disfavor the realization of the twisted superfluid phase.
The phase diagram reported in Fig.\,\ref{fig:06}(b) includes this energy offset $\epsilon$, where $\epsilon>0$ corresponds to the localization of both components on the same sublattice (cf.~Ref.~\cite{Jurgensen:2015}). For $\epsilon\,=\,0$, a continuous degeneracy is predicted between a pure twisted superfluid and a pure spin-density wave, where density waves are anti-aligned for both components [see Fig.\,\ref{fig:06}(a)]. A negative energy offset, for which the two components occupy complementary sublattices, favors such a spin-density wave and destabilizes the twisted superfluid phase. In the opposite case $\epsilon>0$, both components exhibit the same density wave imprinted by the deeper sublattice $\mathbb{A}$ and the twisted superfluid phase remains stable [compare Fig.\,\ref{fig:06}(a)].

In conclusion the twisted superfluid phase can arise in one or two-component systems in presence of pair-tunneling. Its characteristic signature is a symmetry breaking in momentum space. However such a feature can also be induced by scattering during time-of-flight as described in section \ref{sec2}. According to our detailed analysis here and in Ref.~\cite{Jurgensen:2015}, we now assume that the experimental results reported in Ref.~\cite{Soltan-Panahi:2011} are likely due to interactions during the first milliseconds of the time-of-flight expansion. As a central result however, it is possible to evidence the twisted superfluid phase unambiguously by removing one component prior to time-of-flight.

\section{\label{sec4}Kapitza-Dirac diffraction in a state-dependent honeycomb lattice}
In this section we discuss the Kapitza-Dirac diffraction in a state-dependent honeycomb lattice, which induces a similar symmetry breaking in momentum space as discussed before, in this case even for \textit{single component} systems.

When atoms diffract in a standing light field, the atomic momentum can be modified either by zero or two photon momenta as a consequence of two-photon processes (absorption and stimulated emission) \cite{Gould:1986}. Even for far-detuned light, such scattering events might populate higher momenta provided the interaction is sufficiently short and strong.

\begin{figure}[t]
  \centering
  \includegraphics{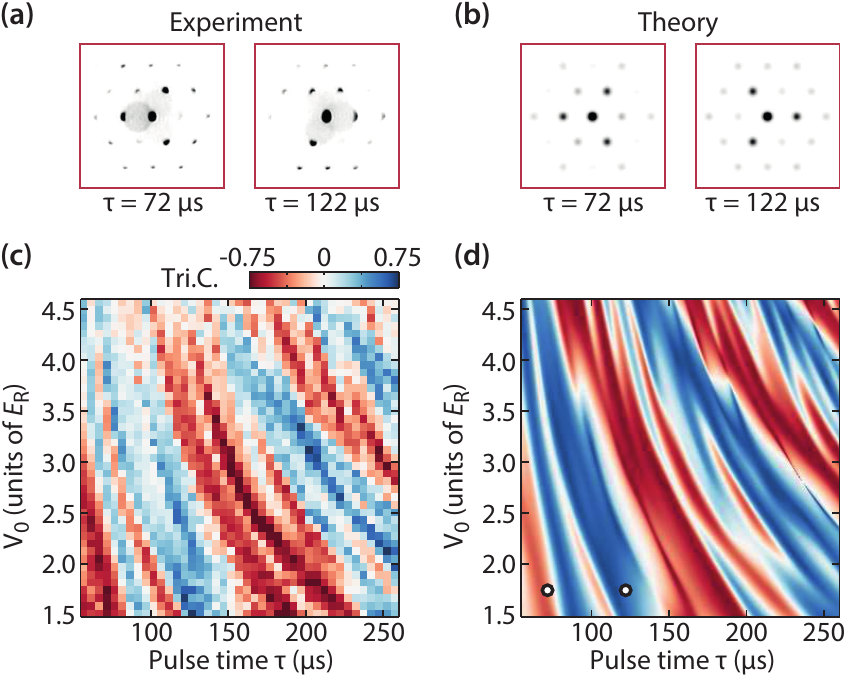}
	\caption[Interaction in TOF]{Kapitza-Dirac diffraction. (a),(b) Measured and calculated diffraction pattern obtained for $\ket{1,-1}$ atoms for two pulse durations $\tau$\,=\,72\,$\mu$s and 122\,$\mu$s at a fixed lattice depth $V_0$\,=\,1.75\,$E_R$. (c),(d) Measured and calculated triangular contrast of the momentum distribution for different pulse duration $\tau$ and lattice depth $V_0$. Circles in (d) mark the positions of the two examples shown in (a) and (b).}
  \label{fig:07}
\end{figure}

The time evolution of the momenta population depends in a striking manner on the hyperfine state of the atoms: for magnetic quantum numbers different from zero, the six fold symmetry of the diffraction pattern breaks, whereas it remains preserved for $m_F\,{=}\,$0. Figure\,\ref{fig:07}(a) shows atomic diffraction patterns obtained by submitting the condensate prepared in the $\ket{1,-1}$ state to the state-dependent honeycomb light field for two different pulse durations $\tau$. The calculated momentum distribution shows an excellent agreement with the experimental data [compare Fig.\,\ref{fig:07}(b)]. For $\ket{1,-1}$ atoms, the triangular contrast has been systematically measured for pulse durations ranging from $50\,\mathrm{\mu s}$ to $250\,\mathrm{\mu s}$ and for different lattice depths $V_0$ as reported in Figure\,\ref{fig:07}(c). For each lattice depth, the triangular contrast oscillates from positive to negative values in a non-periodic manner. This results from the interplay between diffraction in the state-dependent and the state-independent potentials, which have different depths. The experimental data are perfectly reproduced by the numerical simulation of the Schr\"{o}dinger equation reported in Figure\,\ref{fig:07}(d) (see Appendix \ref{AppendixC} for more details).

Indeed, during the interaction with the standing light field, the condensate is subjected to the Hamiltonian:
\begin{equation}
\hat{H}=-\frac{\hbar^2}{2M}(\partial_x^2+\partial_y^2)+V_\text{tot}(x,y).
\end{equation}
Extending the model adopted for one dimensional lattices in \cite{Gadway:2009}, the condensate wave function can be expanded in the basis of plane waves populated by  diffraction $e^{\text{i}2(n\mathbf{b_1}+m\mathbf{b_2})\mathbf{z}}$ with $n$,$m$\,=\,0,$\pm$1,$\pm$2... and $\mathbf{b_1}$, $\mathbf{b_2}$ the two reciprocal lattice vectors (see Appendix \ref{AppendixC}). The numerical simulation of the time evolution of the different momentum states shows an excellent agreement with the measured oscillations of the triangular contrast for all lattice depths as depicted in Figure\,\ref{fig:07}(c).

In conclusion, Kapitza-Dirac diffraction in the state-dependent honeycomb lattice also leads to a strong redistribution of the atoms between different quasi-momenta and induces a symmetry breaking in momentum space for a condensate in hyperfine states with non-zero $g_Fm_F$.

\section{Conclusion}
The results presented in this paper show that the interpretation of time-of-flight measurements as a mapping of the momentum distribution has to be revised for multi-component systems. In particular, scattering of quantum gas mixtures in state-dependent lattices or superlattices can alter the time-of-flight signal substantially. We also demonstrate that an unambiguous identification of the momentum distribution is experimentally possible by removing one component at the initial stage of the time-of-flight expansion. Another possibility is to cancel the inter-species interaction using Feshbach resonances.

Our findings are of central importance for the study of the twisted superfluid phase, which is characterized by a symmetry-broken momentum distribution. We assume that the first observation of a symmetry breaking in the time-of-flight distributions reported in Ref.\,\cite{Soltan-Panahi:2011} can be attributed to interactions during the time-of-flight expansion. Recent theoretical studies \cite{Jurgensen:2015} demonstrate that for larger interaction strengths, the twisted superfluid phase can be realized in a mixture of non-magnetic hyperfine states, for which the energy offset between the two sublattices $\mathbb{A}$ and $\mathbb{B}$ vanishes.

The discussed matter-wave diffraction is not only a limitation for quantum gas experiments relying on time-of-flight information: it can indeed be used as an interferometric probe to reveal novel quantum phases, such as supersolids. A symmetry breaking in momentum space induced by matter-wave diffraction at the start of the time-of-flight expansion could allow for the unambiguous identification of a supersolid.

During the preparation of the manuscript we learned from related work by the Berkeley group of Prof. Stamper-Kurn \cite{StamperKurn2016}.

\section{Acknowledgments}
We thank J.~Struck for valuable comments. This work was supported by the Deutsche Forschungsgemeinschaft under grant SFB 925 and the European Community's Seventh Framework Programme (FP7/2007-2013) under Grant Agreement No. 323714 (EQuaM).

\appendix

\section{Experimental setup}
\label{AppendixA}
In the studies presented here, we work with $^{87}$Rb atoms in different hyperfine states $\ket{F,m_F}$. Radio-frequency and micro-wave couplings of the two hyperfine manifolds $F$\,=\,1,2 allow for the preparation of arbitrary mixtures in the dipole trap.

The state-dependent honeycomb lattice is generated by three linearly polarized laser beams that intersect in the $xy$-plane under angles of $120^\circ$ with the polarization of each laser beam lying in the intersection plane. The quantization axis is defined by a homogeneous magnetic field along the $z$ axis. The alternating circular polarization pattern gives rise to a state-dependent honeycomb potential, which can be written as a sum of a state-independent and a state-dependent part: $V(\mathbf{r})= -V_0[V_{\text{int}}(\mathbf{r}) + V_{\text{pol}}(\mathbf{r})]$.

The state-independent part of the potential reads
\begin{equation}\label{eq:IntPotential}
V_\text{int}(\mathbf{r}) = 6 - 2 \sum_i\cos(\mathbf{b}_i\mathbf{r}),
\end{equation}
where each two reciprocal lattice vectors $\mathbf{b}_i = \varepsilon_{ijk}(\mathbf{k}_j - \mathbf{k}_k)$, given by the corresponding wave vectors of the laser beams $\mathbf{k}_1=2\pi(0,1,0)/\lambda_L$,  $\mathbf{k}_{2/3}=\pi(\pm\sqrt{3},-1,0)/\lambda_L$, span the reciprocal Bravais lattice.

For the quantization axis pointing along the $z$-axis, the state-dependent part of the potential reads
\begin{equation}\label{eq:PolPotential}
    V_\text{pol}(\mathbf{r}) = \sqrt{3}(-1)^{F}m_F\eta \sum_i \sin(\mathbf{b}_i\mathbf{r}),
\end{equation}
where the relative strength of the state-dependent potential is given by the proportionality factor $\eta\,=\,$0.13 that is determined by the detuning of the lattice light with respect to the atomic transitions. The sign change of the Land\'e factor in the ground-state manifold of $^{87}\text{Rb}$ is incorporated in Eq.\,(\ref{eq:PolPotential}) by the prefactor $(-1)^F$.

\section{Numerical simulation of the interaction-induced dynamics in time-of-flight}
\label{AppendixB}
In the following we describe the numerical simulations of the scattering during time-of-flight. We perform time evolution of the Gross-Pitaevskii equation in order to include and evaluate interaction effects during the first milliseconds of time-of-flight.

Our numerical simulation is similar to the calculations in \cite{Pertot:2010}. The authors of \cite{Pertot:2010} prepare their initial state via a Kapitza-Dirac pulse of a one-dimensional lattice. In our setting, the initial state is three-dimensional and features a significant density modulation due to the state-dependent honeycomb optical lattice. Furthermore, we find that the density distribution differs strongly from a Thomas-Fermi profile after time-of-flight, which in turn affects the time evolution. We therefore need to employ a spatially resolved three-dimensional wave function.

The Gross-Pitaevskii equation for the wave function $\Psi^{(\alpha)}(\mathbf{x},t)$ of atomic species $\alpha$ during time-of-flight reads
\begin{equation}\label{eq_gpe}
\begin{split}
&\text{i}\hbar \partial_t \Psi^{(\alpha)}(\mathbf{x},t) = \hat H(t) \Psi^{(\alpha)}(\mathbf{x},t), \text{with} \\
&\hat H(t) = - \frac{\hbar^2}{2 M} \nabla^2 + \sum_\beta g |\Psi^{(\beta)}(\mathbf{x},t)|^2,
\end{split}
\end{equation}
where $g = 4 \pi \hbar^2 a_\mathrm{s}/ M$ is the interaction strength. The s-wave scattering length $a_\mathrm{s} \approx 100 a_0$ is approximated to be spin-independent. We expand the state $\Psi^{(\alpha)}$ in spatially resolved wave functions $\phi_{n,m}(\mathbf{x}, t)$ with momentum $\mathbf{k}_{n,m} = n\mathbf{b}_1 + m\mathbf{b}_2$
\begin{equation}\label{eq:basis}
 \Psi^{(\alpha)}(\mathbf{x},t) = \sum_{n,m} \phi^{(\alpha)}_{n,m} (\mathbf{x} - \frac{\hbar \mathbf{k}_{n,m}}{M}t, t)\ e^{\text{i} \mathbf{k}_{n,m} \mathbf{x}}.
\end{equation}
We restrict the expansion to momenta $|\mathbf{k}_{n,m}| \leq |\mathbf{b}_i|$, since states with larger momenta show negligible occupancy. We find a spatial resolution of $(\Delta x, \Delta y, \Delta z) = (x_0,y_0,z_0)/10$ to be sufficient, as the wave functions $\phi_{n,m}$ only vary slowly.

The initial density distribution is assumed to be a Thomas-Fermi profile, i.e.,
\begin{equation}
 \phi^{(\alpha)}_{n,m}(t=0) = c^{(\alpha)}_{n,m} \sqrt{1 - x^2/x_0^2 - y^2/y_0^2 - z^2/z_0^2}.
\end{equation}
The coefficients $c^{(\alpha)}_{n,m}$ are given by the lowest-band Bloch function at zero momentum, i.e., the solution of a single-particle in an infinite state-dependent honeycomb potential. The extend $\mathbf{r}_0 = (x_0,y_0,z_0)$ of the atomic cloud depends on the trap frequencies and the total particle number $N$.

The time evolution is given by
\begin{equation}\begin{split}
 \phi^{(\alpha)}_{n,m}(\mathbf{x},t + \Delta t)\ =\ \bigg(1 - \frac{\text{i}\hbar |\mathbf{k}_{n,m}|^2}{2 M}\Delta t\bigg) \phi^{(\alpha)}_{n,m}(\mathbf{x},t&) \\
+\ g\!\sum_{\substack{\beta, n',m' \\ n'',m''}} \phi^{(\beta)*}_{n',m'}(\mathbf{x},t)\, \phi^{(\beta)}_{n'',m''}(\mathbf{x},t)\, \phi^{(\alpha)}_{\Delta n,\Delta m}(\mathbf{x},t&),
\end{split}\end{equation}
where $\Delta n = n + n' - n''$ and $\Delta m = m + m' - m''$, due to momentum conservation. The approximation here is that the second derivative of the slowly varying envelope wave functions is negligible and we can assume $\nabla^2 \phi_{n,m}^{(\alpha)} = 0$. For sufficiently small time steps $\Delta t$, the resulting time evolution becomes exact. We find the results to converge for time steps of $\Delta t\,=\,0.01\,\mu s$.

The interaction-induced coupling between the momentum states decays in time as the respective atomic clouds separate spatially. After a separation time of $t_\mathrm{sep} = 2\ \mathrm{max}(x_0, y_0) / (\hbar |\mathbf{b}_i|/M)$ there is no overlap between the clouds anymore. On this timescale, the relative hydrodynamic expansion $\mathrm{max}(x_t/x_0,y_t/y_0,z_t/z_0)$ of the atomic cloud is on the order of $10^{-3}$ and therefore negligible.

\section{Numerical simulations for the Kapitza-Dirac diffraction}
\label{AppendixC}
For the numerical simulations of the Kapitza-Dirac experiment shown in Figure~\ref{fig:07} we perform exact time evolution of a single particle in the state-dependent optical potential. For simplicity, we assume the system to be infinitely large and expand the wave function and potential in terms of plane waves with reciprocal lattice vectors $\mathbf{b}_1$ and  $\mathbf{b}_2$
\begin{equation}\begin{split}
 \Psi(t) &= \sum_{n, m} c_{n, m}(t) e^{\text{i} (n \mathbf{b}_1 + m \mathbf{b}_2)\mathbf{x}}, \\
 V &= \sum_{n, m} d_{n, m} e^{\text{i} (n \mathbf{b}_1 + m \mathbf{b}_2)\mathbf{x}},
\end{split}\end{equation}
with expansion coefficients $c_{n,m}$ and $d_{n,m}$, respectively. The Hamiltonian $\hat H = -\frac{\hbar^2}{2M} \nabla^2 + V$ can be diagonalized exactly with cutoffs in $n$ and $m$. We find plane waves with a momentum $|k| \leq 5 |\mathbf{b}_i|$ to be sufficient for all lattice depths. We assume the initial state to solely occupy zero momentum $c_{n,m}(0) = \delta_{n,0}\delta_{m,0}$.

We use the comparison of the experimental data of the triangular contrast with the theory (Fig. \ref{fig:07}) to calibrate the lattice potential depth and a time offset $t_\mathrm{offset}$ of the pulse length. For the latter, we shift the time axis by $t_\mathrm{offset}$ and find the optimal value for which the deviation between the images is minimal. The obtained correction $t_\mathrm{offset}\,=\,7.3\,\mu s$ compensates for small uncertainties in the experimental procedure.


\begin{thebibliography}{28}%
	\makeatletter
	\providecommand \@ifxundefined [1]{%
		\@ifx{#1\undefined}
	}%
	\providecommand \@ifnum [1]{%
		\ifnum #1\expandafter \@firstoftwo
		\else \expandafter \@secondoftwo
		\fi
	}%
	\providecommand \@ifx [1]{%
		\ifx #1\expandafter \@firstoftwo
		\else \expandafter \@secondoftwo
		\fi
	}%
	\providecommand \natexlab [1]{#1}%
	\providecommand \enquote  [1]{``#1''}%
	\providecommand \bibnamefont  [1]{#1}%
	\providecommand \bibfnamefont [1]{#1}%
	\providecommand \citenamefont [1]{#1}%
	\providecommand \href@noop [0]{\@secondoftwo}%
	\providecommand \href [0]{\begingroup \@sanitize@url \@href}%
	\providecommand \@href[1]{\@@startlink{#1}\@@href}%
	\providecommand \@@href[1]{\endgroup#1\@@endlink}%
	\providecommand \@sanitize@url [0]{\catcode `\\12\catcode `\$12\catcode
		`\&12\catcode `\#12\catcode `\^12\catcode `\_12\catcode `\%12\relax}%
	\providecommand \@@startlink[1]{}%
	\providecommand \@@endlink[0]{}%
	\providecommand \url  [0]{\begingroup\@sanitize@url \@url }%
	\providecommand \@url [1]{\endgroup\@href {#1}{\urlprefix }}%
	\providecommand \urlprefix  [0]{URL }%
	\providecommand \Eprint [0]{\href }%
	\providecommand \doibase [0]{http://dx.doi.org/}%
	\providecommand \selectlanguage [0]{\@gobble}%
	\providecommand \bibinfo  [0]{\@secondoftwo}%
	\providecommand \bibfield  [0]{\@secondoftwo}%
	\providecommand \translation [1]{[#1]}%
	\providecommand \BibitemOpen [0]{}%
	\providecommand \bibitemStop [0]{}%
	\providecommand \bibitemNoStop [0]{.\EOS\space}%
	\providecommand \EOS [0]{\spacefactor3000\relax}%
	\providecommand \BibitemShut  [1]{\csname bibitem#1\endcsname}%
	\let\auto@bib@innerbib\@empty
	\bibitem [{\citenamefont {Bloch}\ \emph {et~al.}(2008)\citenamefont {Bloch},
		\citenamefont {Dalibard},\ and\ \citenamefont {Zwerger}}]{Bloch:2008gl}%
	\BibitemOpen
	\bibfield  {author} {\bibinfo {author} {\bibfnamefont {I.}~\bibnamefont
			{Bloch}}, \bibinfo {author} {\bibfnamefont {J.}~\bibnamefont {Dalibard}}, \
		and\ \bibinfo {author} {\bibfnamefont {W.}~\bibnamefont {Zwerger}},\ }\href
	{\doibase 10.1103/RevModPhys.80.885} {\bibfield  {journal} {\bibinfo
			{journal} {Rev. Mod. Phys.}\ }\textbf {\bibinfo {volume} {80}},\ \bibinfo
		{pages} {885} (\bibinfo {year} {2008})}\BibitemShut {NoStop}%
	\bibitem [{\citenamefont {Lewenstein}\ \emph {et~al.}(2012)\citenamefont
		{Lewenstein}, \citenamefont {Sanpera},\ and\ \citenamefont
		{Ahufinger}}]{Lewenstein:WpY7sZa0}%
	\BibitemOpen
	\bibfield  {author} {\bibinfo {author} {\bibfnamefont {M.}~\bibnamefont
			{Lewenstein}}, \bibinfo {author} {\bibfnamefont {A.}~\bibnamefont {Sanpera}},
		\ and\ \bibinfo {author} {\bibfnamefont {V.}~\bibnamefont {Ahufinger}},\
	}\href {http://ukcatalogue.oup.com/product/9780199573127.do{\#}.UQ1fu2f_baI}
	{\emph {\bibinfo {title} {Ultracold Atoms in Optical Lattices}}}\ (\bibinfo
	{publisher} {Oxford University Press},\ \bibinfo {year} {2012})\BibitemShut
	{NoStop}%
	\bibitem [{\citenamefont {Grynberg}\ \emph {et~al.}(1993)\citenamefont
		{Grynberg}, \citenamefont {Lounis}, \citenamefont {Verkerk}, \citenamefont
		{Courtois},\ and\ \citenamefont {Salomon}}]{Grynberg:1993bm}%
	\BibitemOpen
	\bibfield  {author} {\bibinfo {author} {\bibfnamefont {G.}~\bibnamefont
			{Grynberg}}, \bibinfo {author} {\bibfnamefont {B.}~\bibnamefont {Lounis}},
		\bibinfo {author} {\bibfnamefont {P.}~\bibnamefont {Verkerk}}, \bibinfo
		{author} {\bibfnamefont {J.~Y.}\ \bibnamefont {Courtois}}, \ and\ \bibinfo
		{author} {\bibfnamefont {C.}~\bibnamefont {Salomon}},\ }\href {\doibase
		10.1103/PhysRevLett.70.2249} {\bibfield  {journal} {\bibinfo  {journal}
			{Phys. Rev. Lett.}\ }\textbf {\bibinfo {volume} {70}},\ \bibinfo {pages}
		{2249} (\bibinfo {year} {1993})}\BibitemShut {NoStop}%
	\bibitem [{\citenamefont {Becker}\ \emph {et~al.}(2010)\citenamefont {Becker},
		\citenamefont {Soltan-Panahi}, \citenamefont {Kronj{\"a}ger}, \citenamefont
		{D{\"o}rscher}, \citenamefont {Bongs},\ and\ \citenamefont
		{Sengstock}}]{Becker:2010de}%
	\BibitemOpen
	\bibfield  {author} {\bibinfo {author} {\bibfnamefont {C.}~\bibnamefont
			{Becker}}, \bibinfo {author} {\bibfnamefont {P.}~\bibnamefont
			{Soltan-Panahi}}, \bibinfo {author} {\bibfnamefont {J.}~\bibnamefont
			{Kronj{\"a}ger}}, \bibinfo {author} {\bibfnamefont {S.}~\bibnamefont
			{D{\"o}rscher}}, \bibinfo {author} {\bibfnamefont {K.}~\bibnamefont {Bongs}},
		\ and\ \bibinfo {author} {\bibfnamefont {K.}~\bibnamefont {Sengstock}},\
	}\href {\doibase 10.1088/1367-2630/12/6/065025} {\bibfield  {journal}
	{\bibinfo  {journal} {New J. Phys.}\ }\textbf {\bibinfo {volume} {12}},\
	\bibinfo {pages} {065025} (\bibinfo {year} {2010})}\BibitemShut {NoStop}%
\bibitem [{\citenamefont {Soltan-Panahi}\ \emph
	{et~al.}(2011{\natexlab{a}})\citenamefont {Soltan-Panahi}, \citenamefont
	{Struck}, \citenamefont {Hauke}, \citenamefont {Bick}, \citenamefont
	{Plenkers}, \citenamefont {Meineke}, \citenamefont {Becker}, \citenamefont
	{Windpassinger}, \citenamefont {Lewenstein},\ and\ \citenamefont
	{Sengstock}}]{Soltan-Panahi:2011ey}%
\BibitemOpen
\bibfield  {author} {\bibinfo {author} {\bibfnamefont {P.}~\bibnamefont
		{Soltan-Panahi}}, \bibinfo {author} {\bibfnamefont {J.}~\bibnamefont
		{Struck}}, \bibinfo {author} {\bibfnamefont {P.}~\bibnamefont {Hauke}},
	\bibinfo {author} {\bibfnamefont {A.}~\bibnamefont {Bick}}, \bibinfo {author}
	{\bibfnamefont {W.}~\bibnamefont {Plenkers}}, \bibinfo {author}
	{\bibfnamefont {G.}~\bibnamefont {Meineke}}, \bibinfo {author} {\bibfnamefont
		{C.}~\bibnamefont {Becker}}, \bibinfo {author} {\bibfnamefont
		{P.}~\bibnamefont {Windpassinger}}, \bibinfo {author} {\bibfnamefont
		{M.}~\bibnamefont {Lewenstein}}, \ and\ \bibinfo {author} {\bibfnamefont
		{K.}~\bibnamefont {Sengstock}},\ }\href {\doibase 10.1038/nphys1916}
{\bibfield  {journal} {\bibinfo  {journal} {Nature Phys.}\ }\textbf {\bibinfo
		{volume} {7}},\ \bibinfo {pages} {434} (\bibinfo {year}
	{2011}{\natexlab{a}})}\BibitemShut {NoStop}%
\bibitem [{\citenamefont {Tarruell}\ \emph {et~al.}(2012)\citenamefont
	{Tarruell}, \citenamefont {Greif}, \citenamefont {Uehlinger}, \citenamefont
	{Jotzu},\ and\ \citenamefont {Esslinger}}]{Tarruell:2012db}%
\BibitemOpen
\bibfield  {author} {\bibinfo {author} {\bibfnamefont {L.}~\bibnamefont
		{Tarruell}}, \bibinfo {author} {\bibfnamefont {D.}~\bibnamefont {Greif}},
	\bibinfo {author} {\bibfnamefont {T.}~\bibnamefont {Uehlinger}}, \bibinfo
	{author} {\bibfnamefont {G.}~\bibnamefont {Jotzu}}, \ and\ \bibinfo {author}
	{\bibfnamefont {T.}~\bibnamefont {Esslinger}},\ }\href {\doibase
	10.1038/nature10871} {\bibfield  {journal} {\bibinfo  {journal} {Nature}\
	}\textbf {\bibinfo {volume} {483}},\ \bibinfo {pages} {302} (\bibinfo {year}
	{2012})}\BibitemShut {NoStop}%
\bibitem [{\citenamefont {Jo}\ \emph {et~al.}(2012)\citenamefont {Jo},
	\citenamefont {Guzman}, \citenamefont {Thomas}, \citenamefont {Hosur},
	\citenamefont {Vishwanath},\ and\ \citenamefont {Stamper-Kurn}}]{Jo:2012br}%
\BibitemOpen
\bibfield  {author} {\bibinfo {author} {\bibfnamefont {G.-B.}\ \bibnamefont
		{Jo}}, \bibinfo {author} {\bibfnamefont {J.}~\bibnamefont {Guzman}}, \bibinfo
	{author} {\bibfnamefont {C.~K.}\ \bibnamefont {Thomas}}, \bibinfo {author}
	{\bibfnamefont {P.}~\bibnamefont {Hosur}}, \bibinfo {author} {\bibfnamefont
		{A.}~\bibnamefont {Vishwanath}}, \ and\ \bibinfo {author} {\bibfnamefont
		{D.~M.}\ \bibnamefont {Stamper-Kurn}},\ }\href {\doibase
	10.1103/PhysRevLett.108.045305} {\bibfield  {journal} {\bibinfo  {journal}
		{Phys. Rev. Lett.}\ }\textbf {\bibinfo {volume} {108}},\ \bibinfo {pages}
	{045305} (\bibinfo {year} {2012})}\BibitemShut {NoStop}%
\bibitem [{\citenamefont {Mandel}\ \emph {et~al.}(2003)\citenamefont {Mandel},
	\citenamefont {Greiner}, \citenamefont {Widera}, \citenamefont {Rom},
	\citenamefont {H{\"a}nsch},\ and\ \citenamefont {Bloch}}]{Mandel:2003fj}%
\BibitemOpen
\bibfield  {author} {\bibinfo {author} {\bibfnamefont {O.}~\bibnamefont
		{Mandel}}, \bibinfo {author} {\bibfnamefont {M.}~\bibnamefont {Greiner}},
	\bibinfo {author} {\bibfnamefont {A.}~\bibnamefont {Widera}}, \bibinfo
	{author} {\bibfnamefont {T.}~\bibnamefont {Rom}}, \bibinfo {author}
	{\bibfnamefont {T.}~\bibnamefont {H{\"a}nsch}}, \ and\ \bibinfo {author}
	{\bibfnamefont {I.}~\bibnamefont {Bloch}},\ }\href {\doibase
	10.1038/nature02008} {\bibfield  {journal} {\bibinfo  {journal} {Nature}\
	}\textbf {\bibinfo {volume} {425}},\ \bibinfo {pages} {937} (\bibinfo {year}
	{2003})}\BibitemShut {NoStop}%
\bibitem [{\citenamefont {Karski}\ \emph {et~al.}(2009)\citenamefont {Karski},
	\citenamefont {Förster}, \citenamefont {Choi}, \citenamefont {Steffen},
	\citenamefont {Alt}, \citenamefont {Meschede},\ and\ \citenamefont
	{Widera}}]{Karski:2009}%
\BibitemOpen
\bibfield  {author} {\bibinfo {author} {\bibfnamefont {M.}~\bibnamefont
		{Karski}}, \bibinfo {author} {\bibfnamefont {L.}~\bibnamefont {Förster}},
	\bibinfo {author} {\bibfnamefont {J.-M.}\ \bibnamefont {Choi}}, \bibinfo
	{author} {\bibfnamefont {A.}~\bibnamefont {Steffen}}, \bibinfo {author}
	{\bibfnamefont {W.}~\bibnamefont {Alt}}, \bibinfo {author} {\bibfnamefont
		{D.}~\bibnamefont {Meschede}}, \ and\ \bibinfo {author} {\bibfnamefont
		{A.}~\bibnamefont {Widera}},\ }\href {\doibase 10.1126/science.1174436}
{\bibfield  {journal} {\bibinfo  {journal} {Science}\ }\textbf {\bibinfo
		{volume} {325}},\ \bibinfo {pages} {174} (\bibinfo {year}
	{2009})}\BibitemShut {NoStop}%
\bibitem [{\citenamefont {Lee}\ \emph {et~al.}(2007)\citenamefont {Lee},
	\citenamefont {Anderlini}, \citenamefont {Brown}, \citenamefont
	{Sebby-Strabley}, \citenamefont {Phillips},\ and\ \citenamefont
	{Porto}}]{Lee:2007}%
\BibitemOpen
\bibfield  {author} {\bibinfo {author} {\bibfnamefont {P.~J.}\ \bibnamefont
		{Lee}}, \bibinfo {author} {\bibfnamefont {M.}~\bibnamefont {Anderlini}},
	\bibinfo {author} {\bibfnamefont {B.~L.}\ \bibnamefont {Brown}}, \bibinfo
	{author} {\bibfnamefont {J.}~\bibnamefont {Sebby-Strabley}}, \bibinfo
	{author} {\bibfnamefont {W.~D.}\ \bibnamefont {Phillips}}, \ and\ \bibinfo
	{author} {\bibfnamefont {J.~V.}\ \bibnamefont {Porto}},\ }\href {\doibase
	10.1103/PhysRevLett.99.020402} {\bibfield  {journal} {\bibinfo  {journal}
		{Phys. Rev. Lett.}\ }\textbf {\bibinfo {volume} {99}},\ \bibinfo {pages}
	{020402} (\bibinfo {year} {2007})}\BibitemShut {NoStop}%
\bibitem [{\citenamefont {McKay}\ and\ \citenamefont
	{DeMarco}(2010)}]{McKay:2010jn}%
\BibitemOpen
\bibfield  {author} {\bibinfo {author} {\bibfnamefont {D.}~\bibnamefont
		{McKay}}\ and\ \bibinfo {author} {\bibfnamefont {B.}~\bibnamefont
		{DeMarco}},\ }\href {\doibase 10.1088/1367-2630/12/5/055013} {\bibfield
	{journal} {\bibinfo  {journal} {New J. Phys.}\ }\textbf {\bibinfo {volume}
		{12}},\ \bibinfo {pages} {055013} (\bibinfo {year} {2010})}\BibitemShut
{NoStop}%
\bibitem [{\citenamefont {Bakr}\ \emph {et~al.}(2009)\citenamefont {Bakr},
	\citenamefont {Gillen}, \citenamefont {Peng}, \citenamefont {F{\"o}lling},\
	and\ \citenamefont {Greiner}}]{Bakr:2009bx}%
\BibitemOpen
\bibfield  {author} {\bibinfo {author} {\bibfnamefont {W.}~\bibnamefont
		{Bakr}}, \bibinfo {author} {\bibfnamefont {J.}~\bibnamefont {Gillen}},
	\bibinfo {author} {\bibfnamefont {A.}~\bibnamefont {Peng}}, \bibinfo {author}
	{\bibfnamefont {S.}~\bibnamefont {F{\"o}lling}}, \ and\ \bibinfo {author}
	{\bibfnamefont {M.}~\bibnamefont {Greiner}},\ }\href {\doibase
	10.1038/nature08482} {\bibfield  {journal} {\bibinfo  {journal} {Nature}\
	}\textbf {\bibinfo {volume} {462}},\ \bibinfo {pages} {74} (\bibinfo {year}
	{2009})}\BibitemShut {NoStop}%
\bibitem [{\citenamefont {Sherson}\ \emph {et~al.}(2010)\citenamefont
	{Sherson}, \citenamefont {Weitenberg}, \citenamefont {Endres}, \citenamefont
	{Cheneau}, \citenamefont {Bloch},\ and\ \citenamefont
	{Kuhr}}]{Sherson:2010hg}%
\BibitemOpen
\bibfield  {author} {\bibinfo {author} {\bibfnamefont {J.}~\bibnamefont
		{Sherson}}, \bibinfo {author} {\bibfnamefont {C.}~\bibnamefont {Weitenberg}},
	\bibinfo {author} {\bibfnamefont {M.}~\bibnamefont {Endres}}, \bibinfo
	{author} {\bibfnamefont {M.}~\bibnamefont {Cheneau}}, \bibinfo {author}
	{\bibfnamefont {I.}~\bibnamefont {Bloch}}, \ and\ \bibinfo {author}
	{\bibfnamefont {S.}~\bibnamefont {Kuhr}},\ }\href {\doibase
	10.1038/nature09378} {\bibfield  {journal} {\bibinfo  {journal} {Nature}\
	}\textbf {\bibinfo {volume} {467}},\ \bibinfo {pages} {68} (\bibinfo {year}
	{2010})}\BibitemShut {NoStop}%
\bibitem [{\citenamefont {Haller}\ \emph {et~al.}(2015)\citenamefont {Haller},
	\citenamefont {Hudson}, \citenamefont {Kelly}, \citenamefont {Cotta},
	\citenamefont {Peaudecerf}, \citenamefont {Bruce},\ and\ \citenamefont
	{Kuhr}}]{Haller:2015}%
\BibitemOpen
\bibfield  {author} {\bibinfo {author} {\bibfnamefont {E.}~\bibnamefont
		{Haller}}, \bibinfo {author} {\bibfnamefont {J.}~\bibnamefont {Hudson}},
	\bibinfo {author} {\bibfnamefont {A.}~\bibnamefont {Kelly}}, \bibinfo
	{author} {\bibfnamefont {D.~A.}\ \bibnamefont {Cotta}}, \bibinfo {author}
	{\bibfnamefont {B.}~\bibnamefont {Peaudecerf}}, \bibinfo {author}
	{\bibfnamefont {G.~D.}\ \bibnamefont {Bruce}}, \ and\ \bibinfo {author}
	{\bibfnamefont {S.}~\bibnamefont {Kuhr}},\ }\href {\doibase
	10.1038/nphys3403} {\bibfield  {journal} {\bibinfo  {journal} {Nature
			Physics}\ }\textbf {\bibinfo {volume} {11}},\ \bibinfo {pages} {738}
	(\bibinfo {year} {2015})}\BibitemShut {NoStop}%
\bibitem [{\citenamefont {Cheuk}\ \emph {et~al.}(2015)\citenamefont {Cheuk},
	\citenamefont {Nichols}, \citenamefont {Okan}, \citenamefont {Gersdorf},
	\citenamefont {Ramasesh}, \citenamefont {Bakr}, \citenamefont {Lompe},\ and\
	\citenamefont {Zwierlein}}]{Cheuk:2015}%
\BibitemOpen
\bibfield  {author} {\bibinfo {author} {\bibfnamefont {L.~W.}\ \bibnamefont
		{Cheuk}}, \bibinfo {author} {\bibfnamefont {M.~A.}\ \bibnamefont {Nichols}},
	\bibinfo {author} {\bibfnamefont {M.}~\bibnamefont {Okan}}, \bibinfo {author}
	{\bibfnamefont {T.}~\bibnamefont {Gersdorf}}, \bibinfo {author}
	{\bibfnamefont {V.~V.}\ \bibnamefont {Ramasesh}}, \bibinfo {author}
	{\bibfnamefont {W.~S.}\ \bibnamefont {Bakr}}, \bibinfo {author}
	{\bibfnamefont {T.}~\bibnamefont {Lompe}}, \ and\ \bibinfo {author}
	{\bibfnamefont {M.~W.}\ \bibnamefont {Zwierlein}},\ }\href {\doibase
	10.1103/PhysRevLett.114.193001} {\bibfield  {journal} {\bibinfo  {journal}
		{Phys. Rev. Lett.}\ }\textbf {\bibinfo {volume} {114}},\ \bibinfo {pages}
	{193001} (\bibinfo {year} {2015})}\BibitemShut {NoStop}%
\bibitem [{\citenamefont {Parsons}\ \emph {et~al.}(2015)\citenamefont
	{Parsons}, \citenamefont {Huber}, \citenamefont {Mazurenko}, \citenamefont
	{Chiu}, \citenamefont {Setiawan}, \citenamefont {Wooley-Brown}, \citenamefont
	{Blatt},\ and\ \citenamefont {Greiner}}]{Parsons:2015}%
\BibitemOpen
\bibfield  {author} {\bibinfo {author} {\bibfnamefont {M.~F.}\ \bibnamefont
		{Parsons}}, \bibinfo {author} {\bibfnamefont {F.}~\bibnamefont {Huber}},
	\bibinfo {author} {\bibfnamefont {A.}~\bibnamefont {Mazurenko}}, \bibinfo
	{author} {\bibfnamefont {C.~S.}\ \bibnamefont {Chiu}}, \bibinfo {author}
	{\bibfnamefont {W.}~\bibnamefont {Setiawan}}, \bibinfo {author}
	{\bibfnamefont {K.}~\bibnamefont {Wooley-Brown}}, \bibinfo {author}
	{\bibfnamefont {S.}~\bibnamefont {Blatt}}, \ and\ \bibinfo {author}
	{\bibfnamefont {M.}~\bibnamefont {Greiner}},\ }\href {\doibase
	10.1103/PhysRevLett.114.213002} {\bibfield  {journal} {\bibinfo  {journal}
		{Phys. Rev. Lett.}\ }\textbf {\bibinfo {volume} {114}},\ \bibinfo {pages}
	{213002} (\bibinfo {year} {2015})}\BibitemShut {NoStop}%
\bibitem [{\citenamefont {Greiner}\ \emph {et~al.}(2001)\citenamefont
	{Greiner}, \citenamefont {Bloch}, \citenamefont {Mandel}, \citenamefont
	{H{\"a}nsch},\ and\ \citenamefont {Esslinger}}]{Greiner:2001cm}%
\BibitemOpen
\bibfield  {author} {\bibinfo {author} {\bibfnamefont {M.}~\bibnamefont
		{Greiner}}, \bibinfo {author} {\bibfnamefont {I.}~\bibnamefont {Bloch}},
	\bibinfo {author} {\bibfnamefont {O.}~\bibnamefont {Mandel}}, \bibinfo
	{author} {\bibfnamefont {T.}~\bibnamefont {H{\"a}nsch}}, \ and\ \bibinfo
	{author} {\bibfnamefont {T.}~\bibnamefont {Esslinger}},\ }\href {\doibase
	10.1103/PhysRevLett.87.160405} {\bibfield  {journal} {\bibinfo  {journal}
		{Phys. Rev. Lett.}\ }\textbf {\bibinfo {volume} {87}},\ \bibinfo {pages}
	{160405} (\bibinfo {year} {2001})}\BibitemShut {NoStop}%
\bibitem [{\citenamefont {Pedri}\ \emph {et~al.}(2001)\citenamefont {Pedri},
	\citenamefont {Pitaevskii},\ and\ \citenamefont {Stringari}}]{Pedri:2001jn}%
\BibitemOpen
\bibfield  {author} {\bibinfo {author} {\bibfnamefont {P.}~\bibnamefont
		{Pedri}}, \bibinfo {author} {\bibfnamefont {L.}~\bibnamefont {Pitaevskii}}, \
	and\ \bibinfo {author} {\bibfnamefont {S.}~\bibnamefont {Stringari}},\ }\href
{\doibase 10.1103/PhysRevLett.87.220401} {\bibfield  {journal} {\bibinfo
		{journal} {Phys. Rev. Lett.}\ }\textbf {\bibinfo {volume} {87}},\ \bibinfo
	{pages} {220401} (\bibinfo {year} {2001})}\BibitemShut {NoStop}%
\bibitem [{\citenamefont {Soltan-Panahi}\ \emph
	{et~al.}(2011{\natexlab{b}})\citenamefont {Soltan-Panahi}, \citenamefont
	{L{\"u}hmann}, \citenamefont {Struck}, \citenamefont {Windpassinger},\ and\
	\citenamefont {Sengstock}}]{Soltan-Panahi:2011}%
\BibitemOpen
\bibfield  {author} {\bibinfo {author} {\bibfnamefont {P.}~\bibnamefont
		{Soltan-Panahi}}, \bibinfo {author} {\bibfnamefont {D.-S.}\ \bibnamefont
		{L{\"u}hmann}}, \bibinfo {author} {\bibfnamefont {J.}~\bibnamefont {Struck}},
	\bibinfo {author} {\bibfnamefont {P.}~\bibnamefont {Windpassinger}}, \ and\
	\bibinfo {author} {\bibfnamefont {K.}~\bibnamefont {Sengstock}},\ }\href
{\doibase 10.1038/nphys2128} {\bibfield  {journal} {\bibinfo  {journal}
		{Nature Phys.}\ }\textbf {\bibinfo {volume} {8}},\ \bibinfo {pages} {71}
	(\bibinfo {year} {2011}{\natexlab{b}})}\BibitemShut {NoStop}%
\bibitem [{\citenamefont {Gerbier}\ \emph {et~al.}(2008)\citenamefont
	{Gerbier}, \citenamefont {Trotzky}, \citenamefont {F{\"o}lling},
	\citenamefont {Schnorrberger}, \citenamefont {Thompson}, \citenamefont
	{Widera}, \citenamefont {Bloch}, \citenamefont {Pollet}, \citenamefont
	{Troyer}, \citenamefont {Capogrosso-Sansone}, \citenamefont {von Prokof'ev},\
	and\ \citenamefont {von Svistunov}}]{Gerbier:2008bs}%
\BibitemOpen
\bibfield  {author} {\bibinfo {author} {\bibfnamefont {F.}~\bibnamefont
		{Gerbier}}, \bibinfo {author} {\bibfnamefont {S.}~\bibnamefont {Trotzky}},
	\bibinfo {author} {\bibfnamefont {S.}~\bibnamefont {F{\"o}lling}}, \bibinfo
	{author} {\bibfnamefont {U.}~\bibnamefont {Schnorrberger}}, \bibinfo {author}
	{\bibfnamefont {J.}~\bibnamefont {Thompson}}, \bibinfo {author}
	{\bibfnamefont {A.}~\bibnamefont {Widera}}, \bibinfo {author} {\bibfnamefont
		{I.}~\bibnamefont {Bloch}}, \bibinfo {author} {\bibfnamefont
		{L.}~\bibnamefont {Pollet}}, \bibinfo {author} {\bibfnamefont
		{M.}~\bibnamefont {Troyer}}, \bibinfo {author} {\bibfnamefont
		{B.}~\bibnamefont {Capogrosso-Sansone}}, \bibinfo {author} {\bibfnamefont
		{N.}~\bibnamefont {von Prokof'ev}}, \ and\ \bibinfo {author} {\bibfnamefont
		{B.}~\bibnamefont {von Svistunov}},\ }\href {\doibase
	10.1103/PhysRevLett.101.155303} {\bibfield  {journal} {\bibinfo  {journal}
		{Phys. Rev. Lett.}\ }\textbf {\bibinfo {volume} {101}},\ \bibinfo {pages}
	{155303} (\bibinfo {year} {2008})}\BibitemShut {NoStop}%
\bibitem [{\citenamefont {Pertot}\ \emph {et~al.}(2010)\citenamefont {Pertot},
	\citenamefont {Gadway},\ and\ \citenamefont {Schneble}}]{Pertot:2010}%
\BibitemOpen
\bibfield  {author} {\bibinfo {author} {\bibfnamefont {D.}~\bibnamefont
		{Pertot}}, \bibinfo {author} {\bibfnamefont {B.}~\bibnamefont {Gadway}}, \
	and\ \bibinfo {author} {\bibfnamefont {D.}~\bibnamefont {Schneble}},\ }\href
{\doibase 10.1103/PhysRevLett.104.200402} {\bibfield  {journal} {\bibinfo
		{journal} {Phys. Rev. Lett.}\ }\textbf {\bibinfo {volume} {104}},\ \bibinfo
	{pages} {200402} (\bibinfo {year} {2010})}\BibitemShut {NoStop}%
\bibitem [{\citenamefont {Gould}\ \emph {et~al.}(1986)\citenamefont {Gould},
	\citenamefont {Ruff},\ and\ \citenamefont {Pritchard}}]{Gould:1986}%
\BibitemOpen
\bibfield  {author} {\bibinfo {author} {\bibfnamefont {P.~L.}\ \bibnamefont
		{Gould}}, \bibinfo {author} {\bibfnamefont {G.~A.}\ \bibnamefont {Ruff}}, \
	and\ \bibinfo {author} {\bibfnamefont {D.~E.}\ \bibnamefont {Pritchard}},\
}\href {\doibase 10.1103/PhysRevLett.56.827} {\bibfield  {journal} {\bibinfo
	{journal} {Phys. Rev. Lett.}\ }\textbf {\bibinfo {volume} {56}},\ \bibinfo
{pages} {827} (\bibinfo {year} {1986})}\BibitemShut {NoStop}%
\bibitem [{\citenamefont {L\"uhmann}\ \emph {et~al.}(2014)\citenamefont
	{L\"uhmann}, \citenamefont {J\"urgensen}, \citenamefont {Weinberg},
	\citenamefont {Simonet}, \citenamefont {Soltan-Panahi},\ and\ \citenamefont
	{Sengstock}}]{Luehmann2014hex}%
\BibitemOpen
\bibfield  {author} {\bibinfo {author} {\bibfnamefont {D.-S.}\ \bibnamefont
		{L\"uhmann}}, \bibinfo {author} {\bibfnamefont {O.}~\bibnamefont
		{J\"urgensen}}, \bibinfo {author} {\bibfnamefont {M.}~\bibnamefont
		{Weinberg}}, \bibinfo {author} {\bibfnamefont {J.}~\bibnamefont {Simonet}},
	\bibinfo {author} {\bibfnamefont {P.}~\bibnamefont {Soltan-Panahi}}, \ and\
	\bibinfo {author} {\bibfnamefont {K.}~\bibnamefont {Sengstock}},\ }\href
{\doibase 10.1103/PhysRevA.90.013614} {\bibfield  {journal} {\bibinfo
		{journal} {Phys. Rev. A}\ }\textbf {\bibinfo {volume} {90}},\ \bibinfo
	{pages} {013614} (\bibinfo {year} {2014})}\BibitemShut {NoStop}%
\bibitem [{\citenamefont {J\"{u}gensen}\ \emph {et~al.}(2015)\citenamefont
	{J\"{u}gensen}, \citenamefont {Sengstock},\ and\ \citenamefont
	{L\"{u}hmann}}]{Jurgensen:2015}%
\BibitemOpen
\bibfield  {author} {\bibinfo {author} {\bibfnamefont {O.}~\bibnamefont
		{J\"{u}gensen}}, \bibinfo {author} {\bibfnamefont {K.}~\bibnamefont
		{Sengstock}}, \ and\ \bibinfo {author} {\bibfnamefont {D.-S.}\ \bibnamefont
		{L\"{u}hmann}},\ }\href {\doibase 10.1038/srep12912} {\bibfield  {journal}
	{\bibinfo  {journal} {Sci. Rep.}\ }\textbf {\bibinfo {volume} {5}},\ \bibinfo
	{pages} {12912} (\bibinfo {year} {2015})}\BibitemShut {NoStop}%
\bibitem [{\citenamefont {Choudhury}\ and\ \citenamefont
	{Mueller}(2013)}]{Choudhury:2013cn}%
\BibitemOpen
\bibfield  {author} {\bibinfo {author} {\bibfnamefont {S.}~\bibnamefont
		{Choudhury}}\ and\ \bibinfo {author} {\bibfnamefont {E.~J.}\ \bibnamefont
		{Mueller}},\ }\href {\doibase 10.1103/PhysRevA.87.033621} {\bibfield
	{journal} {\bibinfo  {journal} {Physical Review A}\ }\textbf {\bibinfo
		{volume} {87}},\ \bibinfo {pages} {033621} (\bibinfo {year}
	{2013})}\BibitemShut {NoStop}%
\bibitem [{\citenamefont {Cao}\ \emph {et~al.}(2015)\citenamefont {Cao},
	\citenamefont {Kr\"onke}, \citenamefont {Stockhofe}, \citenamefont {Simonet},
	\citenamefont {Sengstock}, \citenamefont {L\"uhmann},\ and\ \citenamefont
	{Schmelcher}}]{Cao:2015}%
\BibitemOpen
\bibfield  {author} {\bibinfo {author} {\bibfnamefont {L.}~\bibnamefont
		{Cao}}, \bibinfo {author} {\bibfnamefont {S.}~\bibnamefont {Kr\"onke}},
	\bibinfo {author} {\bibfnamefont {J.}~\bibnamefont {Stockhofe}}, \bibinfo
	{author} {\bibfnamefont {J.}~\bibnamefont {Simonet}}, \bibinfo {author}
	{\bibfnamefont {K.}~\bibnamefont {Sengstock}}, \bibinfo {author}
	{\bibfnamefont {D.-S.}\ \bibnamefont {L\"uhmann}}, \ and\ \bibinfo {author}
	{\bibfnamefont {P.}~\bibnamefont {Schmelcher}},\ }\href {\doibase
	10.1103/PhysRevA.91.043639} {\bibfield  {journal} {\bibinfo  {journal} {Phys.
			Rev. A}\ }\textbf {\bibinfo {volume} {91}},\ \bibinfo {pages} {043639}
	(\bibinfo {year} {2015})}\BibitemShut {NoStop}%
\bibitem [{\citenamefont {Gadway}\ \emph {et~al.}(2009)\citenamefont {Gadway},
	\citenamefont {Pertot}, \citenamefont {Reimann}, \citenamefont {Cohen},\ and\
	\citenamefont {Schneble}}]{Gadway:2009}%
\BibitemOpen
\bibfield  {author} {\bibinfo {author} {\bibfnamefont {B.}~\bibnamefont
		{Gadway}}, \bibinfo {author} {\bibfnamefont {D.}~\bibnamefont {Pertot}},
	\bibinfo {author} {\bibfnamefont {R.}~\bibnamefont {Reimann}}, \bibinfo
	{author} {\bibfnamefont {M.~G.}\ \bibnamefont {Cohen}}, \ and\ \bibinfo
	{author} {\bibfnamefont {D.}~\bibnamefont {Schneble}},\ }\href {\doibase
	10.1364/OE.17.019173} {\bibfield  {journal} {\bibinfo  {journal} {Opt.
			Express}\ }\textbf {\bibinfo {volume} {17}},\ \bibinfo {pages} {19173}
	(\bibinfo {year} {2009})}\BibitemShut {NoStop}%
\bibitem [{\citenamefont {Stamper-Kurn}()}]{StamperKurn2016}%
\BibitemOpen
\bibfield  {author} {\bibinfo {author} {\bibfnamefont {D.~M.}\ \bibnamefont
		{Stamper-Kurn}},\ }\href@noop {} {\bibinfo  {journal} {personal
		communication}\ }\BibitemShut {NoStop}%
\end{thebibliography}
\end{document}